\documentclass[fleqn,usenatbib]{mnras}

\usepackage{graphicx}
\usepackage{lscape}
\usepackage{subcaption}
\usepackage{amsmath}
\usepackage{amssymb}

\title[Disk galaxies at $4 \leq z < 7.7$]{Disk-like galaxies at $4 \lesssim z \lesssim 7.7$: JWST/NIRCam morphologies revealed by denoising VAE–GCNN classification}

\author[S.~S.~Mirzoyan]{
S. S. Mirzoyan$^{1,2}$,
A. Avagyan$^{1}$,
V.~G.~Gurzadyan$^{1,2,3}$\thanks{Corresponding author: gurzadyan@yerphi.am}
\\
$^{1}$Center for Cosmology and Astrophysics, Alikhanyan National Laboratory, Yerevan, Armenia\\
$^{2}$Yerevan State University, Yerevan, Armenia\\
$^{3}$SIA, Sapienza University of Rome, Rome, Italy
}

\date{Received, 20XX}
\pubyear{2026}

\begin{document}
\maketitle

\begin{abstract}
Understanding the prevalence of disk-like galaxies at very high redshifts is crucial for constraining the early formation of angular momentum-supported structures. The advent of JWST now permits rest-frame UV and optical morphological studies deep into cosmic epochs where disks have traditionally been considered uncommon. 
We apply an identical denoising VAE~$\rightarrow$~GCNN classification pipeline to multi-filter JWST/NIRCam cutouts in order to obtain homogeneous, morphology-based disk fractions across the sample.
Our approach comprises two steps: (i) a U-Net Variational Autoencoder (VAE) is trained to remove astrophysical and instrumental contaminants while preserving intrinsic morphology, and (ii) a rotation - and reflection - equivariant GCNN classifier is applied to the denoised cutouts to distinguish disk-like galaxies from non-disks.
We determine the fraction of disk-like galaxies as $\simeq 0.34$ for a sample of JWST 100 galaxies over the redshift range $4 \lesssim z \lesssim 7.7$, also in dependence on the galaxy mass range.
Our GCNN-based morphological analysis indicates that disk-like systems constitute a significant fraction of the considered high-redshift population and underscore the importance of such studies for the models of disk formation in the first billion years.
\end{abstract}


\section{Introduction}
\label{sec:intro}
Understanding when ordered, rotationally supported disks first assemble and survive at high redshift requires a uniform morphological census that is robust to surface-brightness dimming, PSF and bandpass variations, and morphology--$k$--corrections \citep{conselice2014evolution, mager2018galaxy, conselice2024epochs}. The advent of \textit{James Webb Space Telescope (JWST)} now enables rest-frame UV–optical morphology measurements deep into epochs where disks were traditionally expected to be rare \citep{Xu,genin2025dawn,Lee,Bor,carreira2026jwst} and refs therein. Recent visual inspections suggest that spiral-like structure may persist to much earlier times than previously inferred from \textit{HST} imaging \citep{kuhn2023jwst}, reinforcing the need for automated, noise-robust methods capable of operating across heterogeneous high-redshift datasets.
Machine-learning techniques have become central to large-scale morphology studies, with both supervised and unsupervised approaches demonstrating strong performance on diverse imaging surveys \citep{pandya2023,fang2026updated}. However, classifier accuracy remains sensitive to noise and domain-shift effects, especially for faint, high-redshift sources. In Paper~I \citep{mirzoyan2025}, we introduced a U-Net Variational Autoencoder (VAE) to obtain morphology-preserving denoised cutouts, and in Paper~II \citep{mirzoyan2026} showed that this approach improves disk/non-disk separability in \textit{JWST} NIRCam observations up to $z \approx 4$.
In this paper, we extend our analysis to a mass-selected \textit{JWST} sample of massive galaxies over $4 \lesssim z \lesssim 7.7$, enabling a statistically meaningful measurement of the disk-like fraction across a fourfold interval in cosmic time. Applying the same denoising-plus-classification framework introduced in Papers~I and II, we quantify the incidence of disk-like systems and examine its dependence on redshift and galaxy mass. The resulting trends are interpreted within the broader physical context of disk formation and survival under high-redshift conditions, including cold accretion, turbulent support, and merger-driven heating \citep{kocjan2024hot, andalman2025origin}, as well as regarding the role of massive black holes \citep{GO,Rees}. This is among the key issues of matter distribution and structure formation and evolution in the late and early Universe \citep{GFC1,GFC2,GFC3} and the linked cosmological tensions \citep{GS,Cap,Dai,DiV,Dai2}.

\section{Dataset}
\label{sec:dataset}
\begin{figure}
 \centering
 \includegraphics[width=\columnwidth]{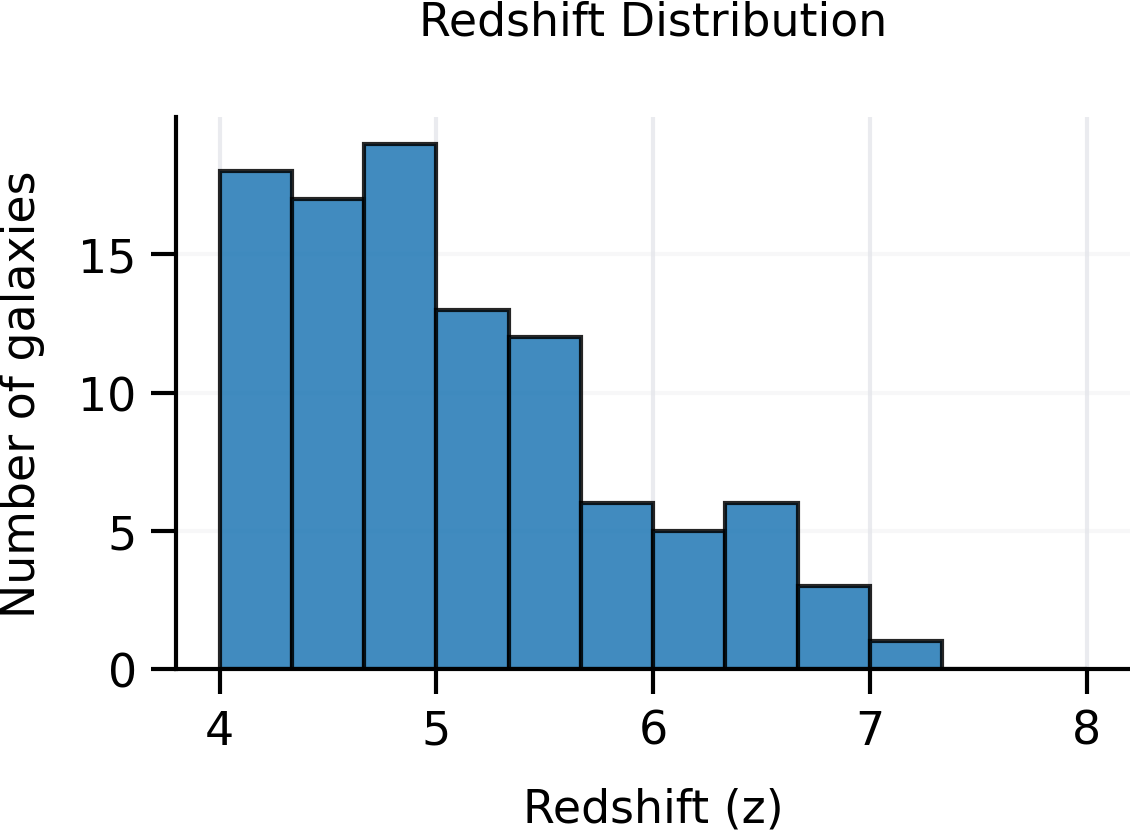}
 \caption{Redshift distribution of the considered JWST galaxy sample.}
 \label{fig:redshift_distribution}
\end{figure}

Our training strategy follows Papers~I \citep{mirzoyan2025} and II \citep{mirzoyan2026}, cf. \citep{M2019}.
The EFIGI catalog \citep{baillard2011efigi} provides, visually inspected galaxies used to train and validate the U--Net VAE denoiser, while Galaxy10\,DECaLS \citep{willett2013galaxy, walmsley2022galaxy} supplies ten-class morphological labels for supervised learning and is employed to train the binary disk/non-disk Group Convolutional Neural Networks (GCNN) classifier.
For the \textit{JWST} sample, we apply a mass-limited and redshift-limited selection motivated by recent CEERS studies of high-$z$ morphology: we retain only CANDELS sources with
$M_\star > 10^{10} M_\odot$ and photometric redshifts in the interval $4 \le z < 7.7$.
These objects are cross-matched to all available NIRCam observations, yielding
1{,}428 cutouts across seven filters (F115W, F150W, F200W, F277W, F356W, F410M, F444W) for 159 unique galaxies. After restricting to the $4 \le z < 7.7$ range and applying the automated image-quality filter to remove extremely low-S/N or structurally incomplete frames, our final high-redshift sample contains 100 galaxies. The distribution of redshifts is presented in Figure~\ref{fig:redshift_distribution}.
As the disk signatures vary strongly with wavelength, each image is classified independently and aggregated at the galaxy level via a multi-instance scheme: a single confident disk detection in any filter is sufficient to label the galaxy as disk-like.

\section{Method}
\label{sec:method}
Our workflow consists of two stages: (i) denoising galaxy cutouts using a U-Net Variational Autoencoder (VAE) trained on realistically contaminated images, and (ii) classifying the cleaned images with a rotation- and reflection-equivariant GCNN to identify disk-like systems.
\subsection{Training data and contamination modeling}
A set of 1,400 visually clean EFIGI galaxies \citep{baillard2011efigi} is used to construct the denoising dataset. Each image is augmented with one to three simulated contaminants---compact or extended sources inserted with randomized profiles, shapes, and orientations---using \texttt{PyAutoLens}, following the contamination strategy adopted in related work \citep{yao2024galaxy}.
We retain 1,000 images for training/validation and 400 for testing, generating ten contamination realizations per clean image for a final corpus of 10,000 denoising examples. All images are padded to $256\times256\times3$ and normalized to $[0,1]$ to ensure consistent VAE training and compatibility with the Galaxy10~DECaLS classification format.
\subsection{U-Net VAE denoiser}
The denoiser is a U-Net–based VAE designed to remove contaminants while preserving faint morphological structure. The encoder includes four convolutional blocks with ReLU activations and max pooling, and the decoder mirrors this structure with transposed convolutions for upsampling.
Training minimizes a standard VAE loss combining binary cross-entropy (BCE) reconstruction and Kullback--Leibler (KL) divergence regularization. The BCE loss is
\begin{equation}
\mathcal{L}_{\mathrm{BCE}}(x,\hat{x})
= -\sum_{i=1}^{n} \left[ x_i \log \hat{x}_i + (1-x_i)\log(1-\hat{x}_i) \right],
\end{equation}
and the KL term is
\begin{equation}
\mathcal{L}_{\mathrm{KL}}
= -\frac{1}{2}\sum_{j=1}^{d}\!\left(1 + \log\sigma_j^{2} - \mu_j^{2} - \sigma_j^{2}\right),
\end{equation}
where $\mu$ and $\sigma$ denote the latent-space mean and standard deviation.
The full objective is
\begin{equation}
\mathcal{L}_{\mathrm{VAE}}
= \mathcal{L}_{\mathrm{BCE}} + \beta\,\mathcal{L}_{\mathrm{KL}},
\end{equation}
with $\beta$ controlling the balance between reconstruction fidelity and latent regularization.
A detailed discussion of model tuning is provided in Paper~II.
\subsection{Binary disk identification with GCNNs}
Denoised images are classified using a dihedral $D_{16}$-equivariant GCNN implemented with the \texttt{escnn} library, following \cite{pandya2023}.
Equivariance to rotations and reflections is crucial for astronomical imaging, where galaxy orientations are arbitrary and high-redshift systems can exhibit diverse morphologies.
Here we focus on a binary disk/non-disk task, extending the approach validated in Paper~II, where the model demonstrated strong performance on Galaxy10~DECaLS and high robustness to noise and geometric variability.
The VAE-cleaned \textit{JWST} images are finally passed through this classifier to produce disk-likelihood scores.

\section{Analysis}
\begin{figure}
 \centering
 \includegraphics[width=\columnwidth]{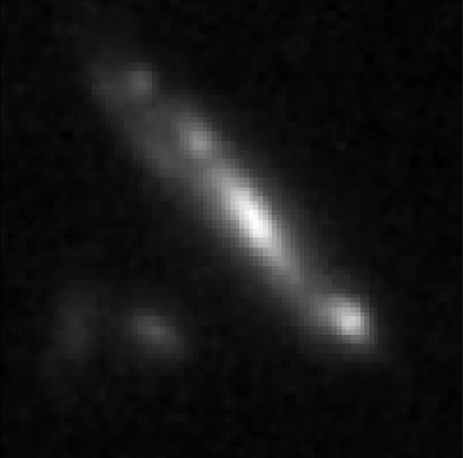}
 \caption{Example of a disk galaxy identified in the F444W filter at a redshift of $z = 5.701$.}
 \label{fig:gal_1}
\end{figure}

To assess the prevalence and demographic trends of disk-like galaxies in our sample, we aggregated the GCNN image level predictions to the galaxy level using the “any positive’’ rule and a noisy OR probability combination scheme. This yielded a catalog of 100 unique galaxies, each having between one and several JWST/NIRCam filter observations. A sample object identified as disk-galaxy is shown in Figure~\ref{fig:gal_1}. Below we present results as functions of redshift, galaxy mass, and observed filter.

\subsection{Redshift dependence}
\begin{figure}
 \centering
 \includegraphics[width=\columnwidth]{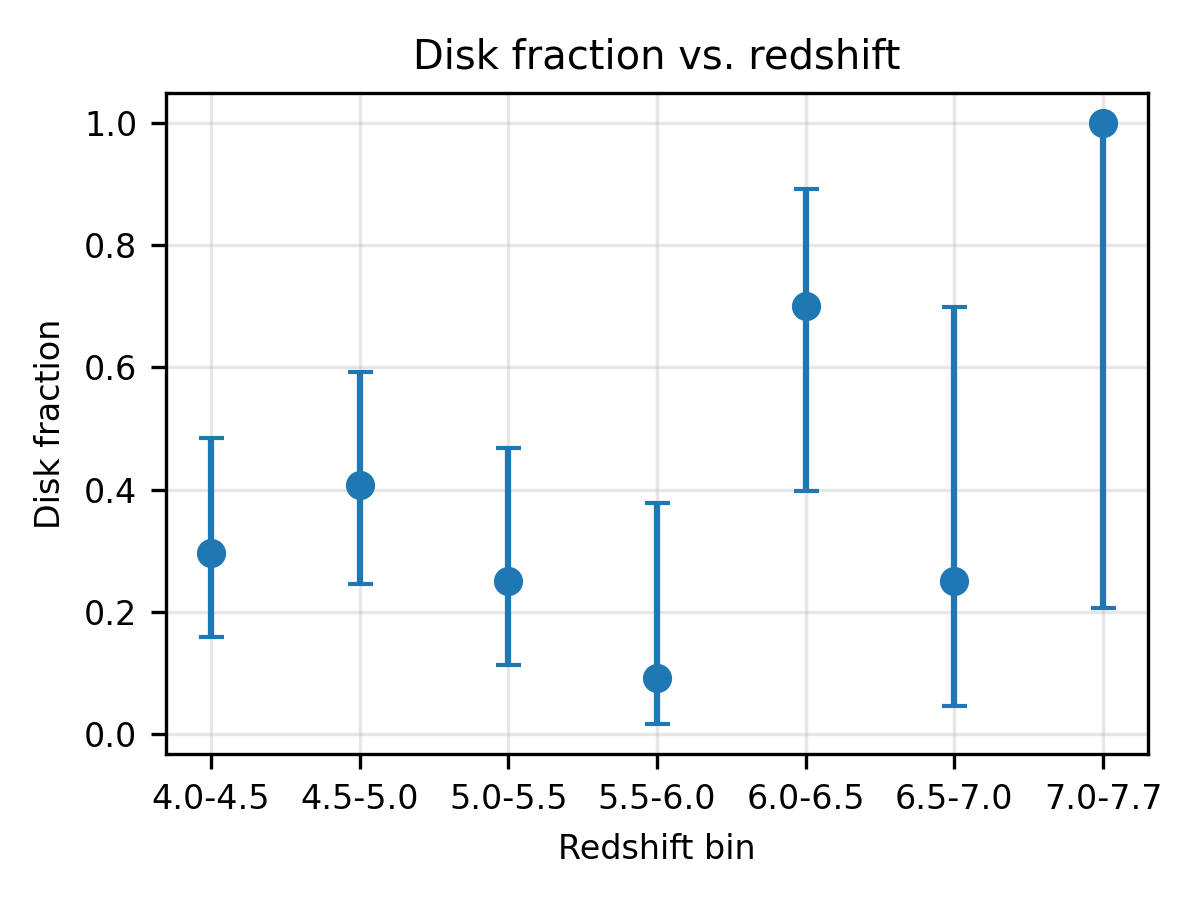}
 \caption{ Disk fraction versus redshift bin with Wilson 95\% confidence intervals.}
 \label{fig:diskfrac_z}
\end{figure}

We binned galaxies in redshift $\Delta z = 0.5$ over the range $4 \leq z < 8$ following our predefined scheme.
We aggregate image-level predictions across available filters to the galaxy level by grouping sources by identifier and applying an “any-positive” rule: as each galaxy is observed in multiple NIRCam filters, we classify images individually and then aggregate them at the galaxy level and a galaxy is labeled disk--like if at least one of its filter images is classified as a disk.
For a continuous confidence metric, we combine per-image probabilities via a noisy-OR aggregator,
\[
p_{\mathrm{gal}} = 1 - \prod_i (1 - p_i),
\]
which matches the physical expectation that disk signatures may be preferentially revealed in specific bands.
The resulting disk fractions, Wilson score intervals, and median galaxy-level probabilities $p_{\rm gal}$ are summarized in Table~\ref{tab:zsummary} and Figure~\ref{fig:diskfrac_z}.
\begin{table}
 \centering
 \caption{Redshift-binned disk fractions (Wilson 95\% CIs).}
 \label{tab:zsummary}
 \vspace{0.25em}
 \begin{tabular}{lrrrrrr}
 $z_{\rm bin}$ & $N$ & $N_{\rm disk}$ & $f_{\rm disk}$ & low & high & $p_{\rm median}$ \\
 4.0--4.5 & 27 & 8 & 0.296 & 0.159 & 0.485 & $\approx 1.0$ \\
 4.5--5.0 & 27 & 11 & 0.407 & 0.245 & 0.593 & $\approx 1.0$ \\
 5.0--5.5 & 20 & 5 & 0.250 & 0.112 & 0.469 & $\approx 1.0$ \\
 5.5--6.0 & 11 & 1 & 0.091 & 0.016 & 0.377 & $\approx 1.0$ \\
 6.0--6.5 & 10 & 7 & 0.700 & 0.397 & 0.892 & $\approx 1.0$ \\
 6.5--7.0 & 4 & 1 & 0.250 & 0.046 & 0.699 & $\approx 1.0$ \\
 7.0--7.7 & 1 & 1 & 1.000 & 0.207 & 1.000 & $\approx 1.0$ \\
 \end{tabular}
\end{table}
At the lowest redshift bin, the disk fraction is $f_{\rm disk}=0.296^{+0.189}_{-0.137}$ with $N=27$ sources. The fraction rises to $0.407^{+0.186}_{-0.162}$ in the $[4.5,5.0)$ bin, again with $N=27$, before declining at $z\sim5$ to $0.250^{+0.219}_{-0.138}$ for $[5.0,5.5)$ (with $N=20$). A local minimum occurs in the $[5.5,6.0)$ interval, where only 1 of 11 galaxies ($f_{\rm disk}=0.091$) is classified as disk-like, albeit with large uncertainties due to small number statistics.
Interestingly, the next bin, $[6.0,6.5)$, shows a pronounced increase: 7 of 10 galaxies are classified as disks, giving $f_{\rm disk}=0.700^{+0.192}_{-0.303}$. Sparse bins at still higher redshift—$[6.5,7.0)$ with $N=4$, and a single object in $[7.0,7.7)$--retain high median probabilities but are not statistically constraining. The median $p_{\rm gal}$ in all bins is effectively unity ($\ge 0.999$), consistent with the high classifier confidence reported at the galaxy level.
Given the modest number of galaxies per bin, the fluctuations may trace the underlying selection function or the sensitivity of the classifier to wavelength-dependent morphological signatures.

\subsection{Disk mass fraction}
We next examined the dependence of disk fraction of galaxy mass using five bins spanning $\log(M_\star/M_\odot)=10.0$--$11.8$
(Table~\ref{tab:msummary}).
\begin{table}
 \centering
 \caption{Disk fraction versus galaxy mass bin with Wilson 95\% confidence intervals.}
 \label{tab:msummary}
 \vspace{0.25em}
 \begin{tabular}{lrrrrr}
 $M_{\rm bin}$ & $N$ & $N_{\rm disk}$ & $f_{\rm disk}$ & low & high \\
 10--10.3 & 62 & 19 & 0.306 & 0.206 & 0.430 \\
 10.3--10.6 & 18 & 8 & 0.444 & 0.246 & 0.663 \\
 10.6--10.9 & 10 & 3 & 0.300 & 0.108 & 0.603 \\
 10.9--11.2 & 5 & 2 & 0.400 & 0.118 & 0.769 \\
 11.2--11.8 & 5 & 2 & 0.400 & 0.118 & 0.769 \\
 \end{tabular}
\end{table}
The lowest mass bin ($10.0$–$10.3$) contains 62 galaxies and exhibits a disk fraction of $f_{\rm disk}=0.306^{+0.124}_{-0.100}$. The next bin ($10.3$--$10.6$) shows a slightly elevated value of $0.444$, based on 8 disks among 18 galaxies, though with broader Wilson uncertainties (0.246--0.663) due to the smaller sample size.
\begin{figure}
 \centering
 \includegraphics[width=\columnwidth]{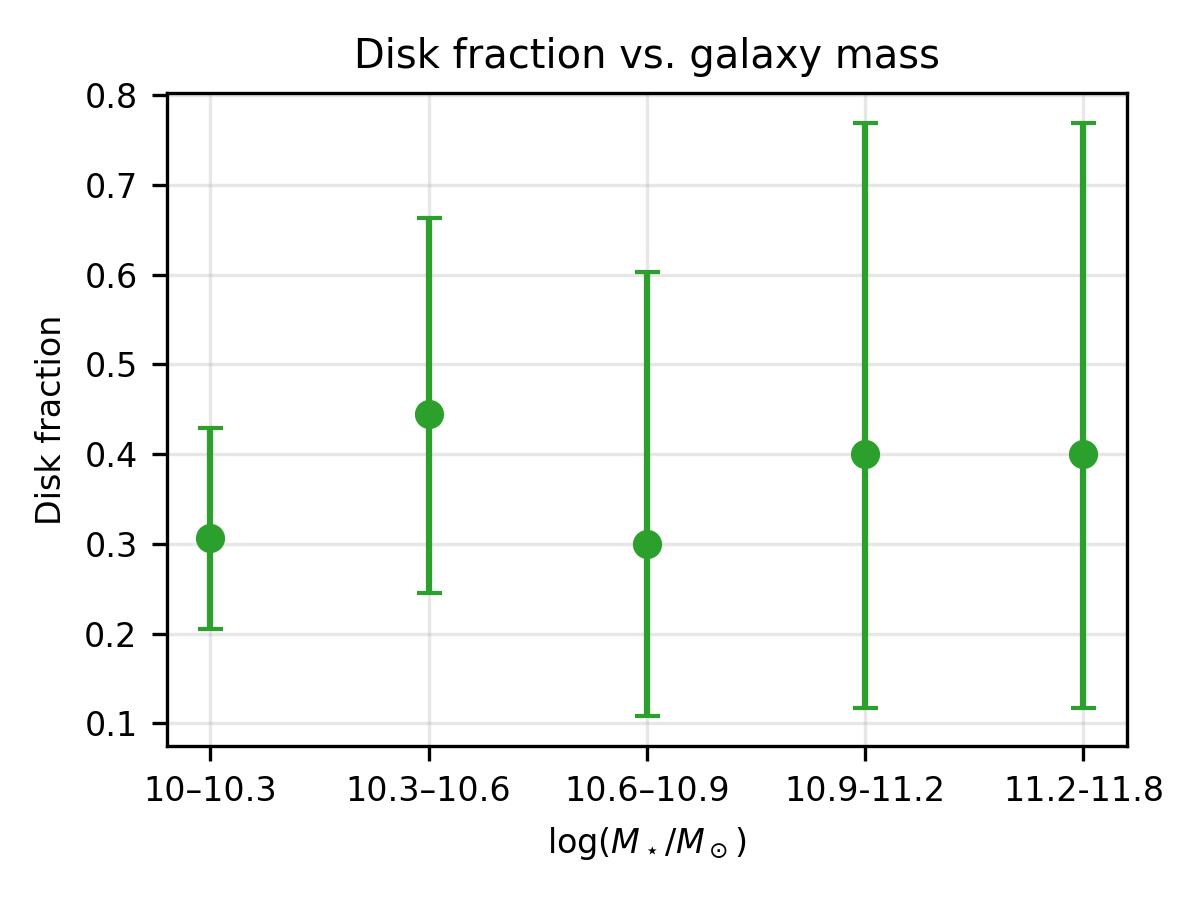}
 \caption{ Disk fraction versus galaxy mass bin with Wilson 95\% confidence intervals.}
 \label{fig:diskfrac_mass}
\end{figure}
Higher mass bins ($10.6$--$10.9$, $10.9$--$11.2$, $11.2$--$11.8$) contain only 10, 5, and 5 galaxies, respectively, and their inferred disk fractions (0.30--0.40) have uncertainties spanning nearly the full 0--1 range (Figure~\ref{fig:diskfrac_mass}).  Within the statistical limitations of the sample, we therefore observe no strong dependence with galaxy mass: the disk fraction remains broadly consistent with $\sim0.3$--$0.4$ across all bins. We used the $\Delta$AIC (Akaike Information Criterion) and $\Delta$BIC (Bayesian Information Criterion) to weight the z-independence for the disk-galaxy fraction and the mass. 
Both, AIC and BIC show slightly greater relative support for a linear trend for the disk-galaxy fraction, than for the
galaxy--mass relation. Specifically, AIC weights $w_{\rm lin} \approx 0.35$ for the z-dependence fraction, compared to 
$w_{\rm lin} \approx 0.30$ for the mass, indicating a marginally higher
preference for a redshift--dependent slope. BIC also favors the z-independence for both cases, again with a
slightly larger value for the disk-fraction ($\mathrm{BF} \approx 0.15$) than for the mass
($\mathrm{BF} \approx 0.12$), reinforcing the same ordering. Taken
together, these diagnostics show that the disk-fraction redshift dependence is only
marginally better supported than the mass' dependence, and both relations
remain fully consistent with redshift independence given the present modest sample size.

\subsection{Filter level behavior and internal consistency}

To quantify bandpass dependent effects, we examined GCNN predictions at the image level for each NIRCam filter, summarized in Table~\ref{tab:fsummary}. The number of images per filter varies from 50 (F150W) to 92 (F356W). The mean predicted probabilities $\langle p\rangle$ are uniformly high (0.898--0.936), reflecting the classifier’s strong internal confidence. Nonetheless, the fraction of images classified as disk-like varies noticeably from filter to filter: F115W shows only 5.6\% disk-positive images, whereas F200W and F444W show higher fractions (13\% and 11.5\%, respectively).
\begin{table}
 \centering
 \caption{Per-filter image diagnostics: number of images ($N$), mean predicted probability ($P_{\rm mean}$), and fraction classified as disk ($f_{\rm disk}$).}
 \label{tab:fsummary}
 \vspace{0.25em}
 \begin{tabular}{lrrr}
 filter & $N$ & $P_{\rm mean}$ & $f_{\rm disk}$ \\
 F115W & 54 & 0.936 & 0.056 \\
 F150W & 50 & 0.898 & 0.060 \\
 F200W & 69 & 0.923 & 0.130 \\
 F277W & 85 & 0.922 & 0.071 \\
 F356W & 92 & 0.911 & 0.098 \\
 F410M & 77 & 0.917 & 0.104 \\
 F444W & 78 & 0.913 & 0.115 \\
 \end{tabular}
\end{table}
These differences, combined with the fact that 32\% of galaxies exhibit mixed labels across their multiple filters (i.e., some positive and some negative classifications), highlight the sensitivity of the classifier to rest-frame wavelength and S/N variations. This filter-level inconsistency is expected for high-redshift targets, where morphological $k$-corrections are substantial and nebular emission can vary strongly across filters.

\section{Conclusions}

We have conducted a GCNN-based morphological analysis of a sample of
100 galaxies observed with JWST/NIRCam, combining multi-filter
image-level predictions into galaxy-level classifications through a
noisy-OR probability scheme. This approach enables a uniform assessment
of disk-like structure across a population spanning $4 \lesssim z
\lesssim 7.7$ and $\log(M_\star/M_\odot)=10$--11.8. The overall disk
fraction of the sample is $f_{\rm disk}\approx0.34$, indicating that
rotationally supported morphologies constitute a significant component
of the early galaxy population.
The analysis, although hints on certain z-dependence for the
disk-galaxy fraction, both it and the masses remain consistent 
with redshift independence and this issue will need further studies with larger samples.

The dependence of disk fraction on the galaxy mass is comparatively weak.
Across all mass bins, the values remain broadly consistent with
$f_{\rm disk}\sim0.3$--$0.4$, with no compelling evidence for a strong
correlation between disk prevalence and galaxy mass within the sampled
range. This suggests that early disk assembly may occur across a broad
mass spectrum, or that mass-related trends are washed out by the modest
sample sizes per bin.
Filter-level behavior reveals systematic variations in image-level
classification, and roughly one-third of galaxies show mixed
disk/non-disk predictions across filters. This underscores the
importance of multi-wavelength morphological diagnostics when
interpreting the structural properties of high-redshift galaxies.
Taken together, these results highlight both the promise and the
complexity of identifying disk-like systems in the early universe.
Future JWST observations with broader wavelength coverage, deeper
exposures, and spectroscopic or kinematic follow-up will be crucial for
disentangling intrinsic structural evolution from observational effects
and for refining our understanding of disk formation at cosmic dawn.

\section*{Acknowledgements}
We are thankful to the referee for valuable comments. We acknowledge the use of the EFIGI, Galaxy10~DECaLS and CEERS JWST databases of https://mast.stsci.edu.

\section*{Data availability}
The data used in this paper are publicly available at https://mast.stsci.edu.


\end{document}